\documentstyle{article}
\def\be{\begin{equation}}
\def\bea{\begin{eqnarray}}
\def\ee{\end{equation}}
\def\eea{\end{eqnarray}}
\def\nn{\nonumber}
\def\bt{\begin{table}}
\def\btab{\begin{tabular}}
\def\et{\end{table}}
\def\etab{\end{tabular}}
\begin{document}
\title {\bf Masses of Heavy Hadrons
}
\author{Harpreet Kaur   \\
Centre for Advanced Study in Physics, Department of Physics, \\
Panjab University, Chandigarh -160014 {\bf India}
}
\maketitle
\begin{abstract}
An estimate has been made of the masses of heavy hadrons in nonrelativistic
quark model, which includes spin and flavor-dependent hyperfine splitting
for two quarks. The effect of variation of the wavefunction value at origin
i.e. $|\psi (0)|^{2}$ and the strong coupling constant $\alpha_{s}$,
with flavor, has also been included in calculating the mass values.
\end{abstract}
\newpage
\section{Introduction}
A lot of data are available on the masses of mesons but the masses of most
baryons have not been measured yet.
Recently predictions about the heavy baryon mass spectrum have become a
subject of increasing interest due to the current experimental
activity of several groups at CERN, Fermilab and the Cornell
Electron Storage Ring (CESR), aimed at the discovery of the baryons
so far absent from the baryon summary table [1]. The copious production of
heavy quarks at LEP, Fermilab Tevatron, CERN LHC and B factories, open for
study the rich spectroscopy of heavy hadrons. So, a plausible theoretical
prediction for the baryon mass spectrum becomes a guide for experimentalists.
\par
Several models [2-10] have been used to evaluate the heavy baryon mass
spectrum.
The nonrelativistic quark model (NRQM), simple and economic one, has been
able to explain very nicely the mass spectrum of light baryons and mesons.
Many workers [3,4] have studied the masses of heavy hadrons
in NRQM. Long time ago, Singh and Khanna [3], ignoring the effect of variation
of strong coupling constant $\alpha_{s}$ and $|\psi (0)|^{2}$
(the wave function value at origin) with flavor, had estimated the masses of
heavy hadrons using the NRQM with
the inclusion of spin and flavor-dependent hyperfine interaction between two
quarks and between a quark and an antiquark. They assume $|\psi (0)|^{2}$
scale for the heavy baryons to be the same as that of hyperons. However,
since $|\psi (0)|^{2}$ is a dimensional quantity it may be incorrect to ignore
its variation with flavor. Evidence to corroborate this has been found in the
quark model [11-13] as well as in the lattice calculation [14]. Moreover,
$\alpha_{s}$ is also scale dependent, so it will be worthwhile to include
its variation with the mass scale.
\par
In this paper, we calculate the masses of heavy hadrons in NRQM with the
inclusion of variation of $|\psi (0)|^{2}$ and $\alpha_{s}$ with flavor.
\section{Mass Formulae}
The mass of a heavy hadron is assumed to arise from the constituent quark
masses plus the two-body hyperfine interaction energies for a meson and a
baryon. Without taking into consideration the variation of $|\psi (0)|^{2}$
and $\alpha_{s}$ with flavor, hadron masses, thus, can be expressed as [3],
\bea
M_{\rm meson}&=&a_{m}+m_{i}+m_{j}+bm_{0}^{2}\frac{{\bf s_{i}.s_{j}}}{m_{i}
m_{j}},
\label{1} \\
M_{\rm baryon}&=&a_{b}+\sum_{i}m_{i}+\frac{bm_{0}^{2}}{3}
\sum_{i<j}\frac{{\bf s_{i}.s_{j}}}{m_{i}m_{j}},
\label{2}
\eea
where $a_{m}$ and $a_{b}$ are the parameters with the dimension of mass and
$m_{i}$ ($m_{j}$) are the masses of respective quarks (antiquarks). $m_{0}$
is a scale factor which is taken to be the mass of the lightest quark i.e.
$m_{0}=m_{u}$. `b', having the dimensions of mass, is a parameter which
includes
the long distance effects i.e. it takes care of the value of the wavefunction
at origin, $|\psi (0)|^{2}$, along with the value of strong coupling constant
$\alpha_{s}$ and is usually taken to be the same for all hadrons,
irrespective of the quark flavor.
\par
Now, if the dependence of $|\psi (0)|^{2}$ and $\alpha_{s}$ on flavor is
taken into account, eqs. (1) and (2) modify to
\bea
M_{\rm meson}&=&a_{m}+m_{i}+m_{j}+b_{ij}m_{0}^{2}\frac{{\bf s_{i}.s_{j}}}{m_{i}
m_{j}},
\label{3} \\
M_{\rm baryon}&=&a_{b}+\sum_{i}m_{i}+\frac{m_{0}^{2}}{3}\sum_{i<j}
b_{ij}\frac{{\bf s_{i}.s_{j}}}{m_{i}m_{j}}.
\label{4}
\eea
Here, the scale factor, $b_{ij}$, instead of having a constant value for all
flavors, is a variable and thus, has different values for different quark
pairs. Hence, $b_{ij}$ is the parameter which takes into consideration the
flavor dependence of $|\psi (0)|^{2}$ and $\alpha_{s}$, so that one has,
\be
b_{ij} \propto |\psi_{ij}(0)|^{2}~\alpha_{s}(\mu).
\label{}
\ee
The value of $b_{ij}$ is taken to be same in the light quark sector i.e.
for any of the u, d and s quark pairs and it varies as one goes to c-sector
and then to b-sector.

\section{Estimation of $|\psi(0)|^{2}$ and $\alpha_{s}(\mu)$}
\subsection{$|\psi(0)|^{2}$}
The flavor dependence reflected in the scale factor $b_{ij}$ corresponding
to the spatial matrix element, is due to the long distance QCD effects.
Evaluation of $|\psi(0)|^{2}$ is as yet uncertain for baryons and more
complicated, because unlike the mesons, these are three-body systems. Infact,
the heavy baryons may provide a good way of testing the flavor dependence of
the confinement forces. The absence of an exact dynamical theory of low energy
interactions between quarks limits our evaluation of $|\psi(0)|^{2}$ from
first principles. However, a naive estimate for the scale parameter may be
obtained using the hyperfine splitting (hfs) [2],
\bea
\Delta E_{\rm hfs} &=& \frac{16\pi \alpha_{s}}{9m_{i}m_{j}}|\psi(0)|^{2}
{\bf s_{i}.s_{j}},
\label{5}
\eea
The experimental hyperfine splittings thus, may provide a reliable
measure of the wavefunction at origin of the ground state baryons
the value of which is needed in the lifetime estimates.
\par
Many workers have tried to estimate the value of $|\psi(0)|^{2}$ for
different quark flavors using different techniques [13-18]. H. Y.
Cheng [15], using formula (5) has found the values respectively, for
c and b sectors to be
\bea
|\psi_{cu}(0)|^{2} &=& 1.5 \times 10^{-2} ~{\rm GeV}^{3}
\label{6}  \\
{\rm and}  \nn \\
|\psi_{bu}(0)|^{2} &=& 2.5 \times 10^{-2} ~{\rm GeV}^{3}.
\label{7}
\eea
K\"orner and Siebert [16], from a fit to hyperfine splitting, have
estimated the value of $|\psi(0)|^{2}$ for c-sector to be nearly
$1.0 \times 10^{-2}$ GeV$^{3}$. As mentioned in ref. [16], NRQM with
a funnel type potential [17] and electromagnetic mass differences in the
hyperfine splitting formula [18] also predict the values similar to
that found by them. Values of the scale parameter $|\psi(0)|^{2}$
for different quark pairs have also been estimated in lattice
approach and have been quoted in ref. [14].
\par
Uppal and Verma [13], using eqn. (5) have made an estimate of the ratio
$\frac{|\psi(0)|_{c}^{2}}{|\psi(0)|_{s}^{2}}$, which comes out to
be approximately 2.83. We have also tried to estimate the wavefunction value
at origin in terms of the ratios, $|\psi_{cu}|^{2}/|\psi_{su}|^{2}$ and
$|\psi_{bu}|^{2}/|\psi_{su}|^{2}$, respectively, for c and b sectors.

\subsection{$\alpha_{s}(\mu)$}
The QCD coupling constant $\alpha_{s}(\mu)$ at any renormalization
scale can be calculated from $\alpha_{s}(m_{Z}) = 0.117$ via
\bea
\alpha_{s}(\mu) &=& \frac{\alpha_{s}(m_{Z})}
{1-(11-\frac{2}{3}n_{f})[\alpha_{s}(m_{Z})/2\pi]ln(m_{Z}/\mu)} ,
\label{}
\eea
and one has,
\be
\alpha_{s}(m_{c}) = 0.31 ~~{\rm and}
~~\alpha_{s}(m_{b}) = 0.20
\label{}
\ee

\section{Numerical Estimate for Masses}
\subsection{Heavy Baryons}
Choosing the set of parameters (in MeV) to be $a_{b}=205, ~m_{0}=m_{u}
=m_{d}=310, ~m_{s}=450, ~m_{c}=1620, ~m_{b}=4976,
~b_{qq^{\prime}}=640, ~b_{qc}=736$ and $b_{qb}=1600$, with $q ~{\rm and}
~q^{\prime} =u, ~d, ~s$,
the masses of heavy baryons are calculated. Following the usual convention
that a particle symbol represents its mass, the
results for c-sector are displayed
in Table 1 whereas Table 2 contains b-baryon masses.
The mass values are in good agreement with the experimental values,
wherever available. For comparison, masses estimated by other techniques
have also been included in respective Tables 1 and 2.
\par
Predictions have also been made for the masses of doubly heavy baryons using
$|\psi_{qc}(0)|^{2} = |\psi_{cc}(0)|^{2}$ and $|\psi_{qb}(0)|^{2} =
|\psi_{hb}(0)|^{2}$, with $h=b$ or $c$. Although it may not be an
appropriate
assumption, as no data are available in this sector, we can still make this
choice. Moreover, when a particle contains two heavy quarks, in the heavy
quark mass limit, the corresponding hyperfine splitting term ($\propto
~1/m_{i}m_{j}$) will give negligible contribution to the overall mass
and hence will be less significant. Predictions made on this assumption
along with those made by other workers, have been given in Tables 1-3.

\subsection{Heavy Mesons}
Many different sets [8,19-29] of constituent quark masses to evaluate the
masses of heavy mesons, are used in the
literature, most of them obtained from fits to spectroscopic data. The values
(in MeV) of various parameters used by us are: $a_{m}=82, ~m_{0}=m_{u}
=m_{d}=310, ~m_{s}=400, ~m_{c}=1578, ~m_{b}=4920, ~b_{qq^{\prime}}
=645, ~b_{qc}=684$ and $b_{qb}=755$, with $q ~{\rm and} ~q^{\prime}=u, ~d,
~s$. Mass values thus obtained are displayed in Table 4 and are in
nice agreement with the experimental values.
\par
It is interesting to note that the quark masses which give a best fit to
the baryons are a little
higher (except for the mass of u quark) than those which lead to
a best fit to the mesons. Because these quark masses are constituent
quark masses, there are no theoretical reasons why the masses determined
from the baryons should coincide exactly with those determined from the
mesons. If we insist that a single set of quark masses hold for both
baryons and mesons, and vary these masses, the overall best fit to the
hadron data is significantly poorer and our predictions have greater
errors.

\section{Discussion}
At present, many of the charm baryon masses are known experimentally,
whereas in b-sector, only the mass of one b-baryon is known accurately.
The charm baryon masses (in MeV) measured till date are [1]
\bea
\Lambda_{c} &=& 2285, \nn \\
\Sigma_{c} &=& 2453 \pm 1, \nn\\
\Xi_{c} &=& 2468 \pm 2, \nn \\
\Omega_{c} &=& 2704 \pm 4, \nn \\
\Sigma_{c}^{*} &=& 2521 \pm 4, \nn\\
\Xi_{c}^{*} &=& 2644 \pm 2.
\label{}
\eea
\par
In 1994, WA89 Collaboration [30] had reported the observation of
$\Xi_{c}^{\prime} = 2563 \pm 15$ MeV. Recent calculations by many
workers [8-10,31-34] consistently predict the mass of
$\Xi_{c}^{\prime}$ to be around 2580 MeV. Also, the preliminary
CLEO results on the $\Xi_{c}^{\prime}-\Xi_{c}$ mass difference
from $\Xi_{c}^{+ \prime} \rightarrow \Xi_{c}^{+} \gamma$,
$\Xi_{c}^{0 \prime} \rightarrow \Xi_{c}^{0} \gamma$ decay [35] give us:
\bea
\Xi_{c}^{+ \prime}-\Xi_{c}^{+} &=& 107.8 \pm 1.7 \pm 2.5 {\rm MeV},
\nn \\
\Xi_{c}^{0 \prime}-\Xi_{c}^{0} &=& 107.0 \pm 1.4 \pm 2.5 {\rm MeV}.
\label{}
\eea
With $\Xi_{c} = 2468 \pm 2$ MeV, these results lead to
\be
\Xi_{c}^{\prime} = 2575 \pm 5 {\rm MeV}.
\label{}
\ee
Our prediction for $\Xi_{c}^{\prime}$ (2580 MeV) is in nice agreement
with this
experimental observation as well as with the predictions made by
other authors. Also the mass difference $\Xi_{c}^{\prime}-\Xi_{c}$,
using our mass values, comes out to be 107 MeV, which coincides very
well with the experimentally measured value.
\par
The $\Omega_{c}^{*}$ as well as double and triple charm baryons,
have not yet been observed. Most of the theoretical calculations
[7-10,31-34,36] predict the mass of $\Omega_{c}^{*}$ to be around 2770 MeV,
whereas predictions for the doubly and triply charmed baryon
masses are less definite. The mass value, as predicted by us, for
$\Omega_{c}^{*}$ (2766 MeV) is consistent with the above mentioned
theoretical predictions.
\par
On the other hand, in the b-sector, only the mass of $\Lambda_{b}$ [1]
baryon is known well and masses of the rest of the b-baryons
are yet to be measured. The mass of $\Lambda_{b}$ baryon as predicted by
us is in nice agreement with the experimental value [1], whereas the
rest of b-baryon masses predicted by us agree well with the predictions
made by ref. [8], as can be seen from Table 2.
\par
To have an estimate of the wavefunction value at origin, we make use
of eqn. (5) i.e. $b_{ij}\propto |\psi_{ij}(0)|^{2}~\alpha_{s}(\mu)$, so that
one has
\bea
\frac{b_{cu}}{b_{su}}& = & \frac{|\psi_{cu}(0)|^{2}\alpha_{s}(m_{c})}
{|\psi_{su}(0)|^{2}\alpha_{s}(m_{s})}
\label{8}
\eea
and
\bea
\frac{b_{bu}}{b_{su}}& = & \frac{|\psi_{bu}(0)|^{2}\alpha_{s}(m_{b})}
{|\psi_{su}(0)|^{2}\alpha_{s}(m_{s})}
\label{9}
\eea
Using our preferred set of parameters, these ratios turn out to be
\bea
\frac{b_{cu}}{b_{su}}&=&1.15
\label{10}
\eea
and
\bea
\frac{b_{bu}}{b_{su}}&=&2.5
\label{11}
\eea
As the mass dependence of the strong coupling constant (eqn. 9)
is known, an estimate of the ratios of the wave function
values at origin (i.e. $|\psi_{cu}|^{2}/|\psi_{su}|^{2}$ and
$|\psi_{bu}|^{2}/|\psi_{su}|^{2}$) can be made using eqs. (15) to (18):
\bea
|\psi_{cu}|^{2}/|\psi_{su}|^{2} &=& 3.3 \nn \\
{\rm and}
~~|\psi_{bu}|^{2}/|\psi_{su}|^{2} &=& 11.3
\label{}
\eea
The ratio for c-sector is little higher than the estimates of ref. [13] and
that of lattice calculations [14]. It may be due to the fact that the ratio
also depends on the strong coupling constant for light sector, whose value
is not fixed. So different values of $\alpha_{s}(m_{s})$ will lead to
different values of these ratios. Only the experimental data in future
will give us better insight in this arena.
\par
Note that, irrespective of introducing the parameters concerning the variation
of wavefunction at origin with flavor, certain independent relations and
mass shifts introduced
in ref. [3] are satisfied here also:
\bea
\Omega_{b}^{*}+\Sigma_{b}^{*}-2\Xi_{b}^{*}=\Omega_{c}^{*}+\Sigma_{c}^{*}
-2\Xi_{c}^{*}
=\Omega^{*}+\Sigma^{*}-2\Xi^{*}, \label{12} \\
(\Sigma_{cc}^{*}-\Sigma_{c}^{*})=\Xi_{cc}-\Sigma_{c},  \label{13} \\
(\Sigma_{bb}^{*}-\Sigma_{b}^{*})=\Xi_{bb}-\Sigma_{b},  \label{14} \\
\frac{(\Sigma_{b}^{*}-\Sigma_{b})}{(B^{*}-B)}=
\frac{(\Sigma_{c}^{*}-\Sigma_{c})}{(D^{*}-D)}=
\frac{(\Sigma^{*}-\Sigma)}{(K^{*}-K)}=\frac{\Delta - N}{\rho - \pi}
=\frac{1}{2} \label{15} \\
\frac{3}{2}\frac{\Sigma_{b}-\Lambda_{b}}{\Delta - N}+\frac{B^{*}-B}
{\rho -\pi}
=\frac{3}{2}\frac{\Sigma_{c}-\Lambda_{c}}{\Delta - N}+\frac{D^{*}-D}
{\rho -\pi}
=\frac{3}{2}\frac{\Sigma-\Lambda}{\Delta - N}+\frac{K^{*}-K}{\rho - \pi}=1
\label{16}
\eea
The predicted masses may have large errors, so it may
be useful to
have an idea of the mass difference from the lowest lying heavy
baryon. In b-sector, one has (in MeV):
\bea
\Sigma_{b} - \Lambda_{b} = 180 , \nn \\
\Sigma_{b}^{*} - \Lambda_{b} = 230 , \nn \\
\Xi_{b} - \Lambda_{b} = 188 ,  \nn \\
\Xi_{b}^{\prime} - \Lambda_{b} = 302 ,  \nn \\
\Xi_{b}^{*} - \Lambda_{b} = 343 , \nn \\
\Omega_{b} - \Lambda_{b} = 436 , \nn \\
\Omega_{b}^{*} - \Lambda_{b} = 469
\label{}
\eea
To conclude, by allowing the variation of parameter $b$ with flavor and
introducing two more parameters, we are able to bring lot of masses of both
baryons and mesons closer to the observed values.

\section{Acknowledgement}
HK thanks Prof. M. P. Khanna for useful discussions and
Council of Scientific and Industrial Research, New Delhi, India, for
fellowship.

\newpage
\begin{table}
\begin{center}
\caption{Masses (in MeV) of  charm baryons}
\vskip 0.4 cm
\begin{tabular}{|c|c|c|c|c|c|}
\hline
Particle&Calculated Mass &Experimental Value &Ref. [3] &Ref. [8] &Ref. [34] \\
\hline 
$\Lambda_{c}$&2285 &2285&2267 & 2285$\pm$1 &-\\
$\Sigma_{c}$&2451 &2455 &2436 &2453$\pm$3 &-\\
$\Xi_{c}$&2475 &2470 &2504  &2468$\pm$3 &-\\
$\Xi_{c}^{\prime}$ &2582 &2580$^{*}$ & 2602 &2580$\pm$20 &2569$\pm$6 \\
$\Omega_{c}$&2717 &2704 & 2775 &2710$\pm$30 &- \\
$\Sigma{c}^{*}$&2521  &2521 & 2491 &2520$\pm$20 &- \\
$\Xi_{c}^{*}$&2642 &2644 &2646 &2650$\pm$20 &- \\
$\Omega_{c}^{*}$&2766 &- &2800 &2770$\pm$30 &2767$\pm$7 \\
$\Xi_{cc}$&3714 &- &3710 &3660$\pm$70 &3610$\pm$3  \\
$\Xi_{cc}^{*}$&3786 &- &3781  &3740$\pm$70 &3735$\pm$17  \\
$\Omega_{cc}$&3882 &- &3865 &3740$\pm$80 &3804$\pm$8  \\
$\Omega_{cc}^{*}$&3914 &- &3948 &3820$\pm$80 &3850$\pm$25 \\
\hline
\end{tabular}
\end{center}
\label{table1}
\end{table}
\begin{table}
\begin{center}
\caption{Masses (in MeV) of beauty baryons}
\vskip 0.4 cm
\begin{tabular}{|c|c|c|c|c|}
\hline
Particle&Calculated Mass &Experimental Value &Ref. [3]&Ref. [8] \\
\hline 
$\Lambda_{b}$ &5641 &5641 &5724 &5620$\pm$40 \\
$\Sigma_{b}$ & 5821 &5814$^{*}$ &5729 & 5820$\pm$40  \\
$\Xi_{b}$&5831 &- &5783 &5810$\pm$40   \\
$\Xi_{b}^{\prime}$ &5949 &- &5893 &5950$\pm$40\\
$\Omega_{b}$&6083 &- &6065 &6060$\pm$50  \\
$\Sigma{b}^{*}$&5871  &5870$^{*}$ &5750 &5850$\pm$40 \\
$\Xi_{b}^{*}$&5991 &- &5915 &5980$\pm$40 \\
$\Omega_{b}^{*}$&6117 &- & 6079&6090$\pm$50  \\
$\Xi_{bb}$&10434 &- &10315 &10340$\pm$100 \\
$\Xi_{bb}^{*}$&10484 &- &10321 &10370$\pm$100  \\
$\Omega_{bb}$&10585 & - &10494 &10370$\pm$100  \\
$\Omega_{bb}$&10619 &- &10321 &10400$\pm$100  \\
\hline
\end{tabular}
\end{center}
$^{*}$ as calculated from the mass differences.
\label{table2}
\end{table}
\newpage
\begin{table}
\begin{center}
\caption{Masses (in MeV) of baryons containing two different heavy quarks
i.e. c and b.}
\vskip 0.4 cm
\begin{tabular}{|c|c|c|c|c|c|c|}
\hline
Particle & $\Xi_{bc}$ & $\Xi_{bc}^{\prime}$ & $\Xi_{bc}^{*}$&$\Omega_{bc}$
&$\Omega_{bc}^{\prime}$&$\Omega_{bc}^{*}$ \\
\hline
Mass&7076&7102&7133&7227&7244&7266 \\
\hline
Ref. [8]&6990$\pm$90 &7040$\pm$90 &7060$\pm$90 &7060$\pm$90 &7090$\pm$90
&7120$\pm$90 \\
\hline
\end{tabular}
\end{center}
\label{table3}
\end{table}
\begin{table}
\begin{center}
\caption{Masses (in MeV) of heavy mesons.}
\vskip 0.4 cm
\begin{tabular}{|c|c|c|}
\hline
Particle&Calculated Mass &Experimental Value [1]  \\
\hline 
$D$ &1869 &1869  \\
$D^{*}$ & 2003 &2007    \\
$D_{s}$&1982 &1968      \\
$D_{s}^{*}$ &2086 &2112 \\
$B$&5276  &5279$\pm$1.8 \\
$B^{*}$&5324 &5325$\pm$1.8 \\
$B_{s}$&5374 &5369$\pm$2  \\
$B_{s}^{*}$&5411&- \\
\hline
\end{tabular}
\end{center}
\label{table4}
\end{table}
\end{document}